\documentclass[preprint,showpacs,preprintnumbers,aps,prb]{revtex4}
\usepackage{amssymb,amsmath}
\usepackage{graphicx}
\usepackage[dvips]{color}
\usepackage{dcolumn}
\usepackage{epstopdf}
\usepackage{epsfig}
\usepackage{bm}

\begin{document}

\title{Origins of dihydrogen binding to metal-inserted porphyrins: electric polarization and Kubas interaction}

\author{Junga Ryou}
\author{Gunn Kim$^{*}$}
\author{Suklyun Hong\footnote{Corresponding authors: hong@sejong.ac.kr; gunnkim@sejong.ac.kr}}
\affiliation{Department of Physics, Graphene Research Institute and Institute of Fundamental Physics, Sejong University, Seoul 143-747, Korea}
\date{\today}

\begin{abstract}
Using density functional theory calculations, we have investigated the interactions between hydrogen molecules and metalloporphyrins. 
A metal atom such as Ca or Ti is introduced for incorporation in the central N$_4$ cavity. 
Within local density approximation (generalized gradient approximation), we find that the average binding energy of H$_2$ to the Ca atom is about 0.25 (0.1) eV/H$_2$ 
up to four H$_2$ molecules, whereas that to the Ti atom is about 0.6 (0.3) eV per H$_2$ up to two H$_2$. 
Our analysis of orbital hybridization between the inserted metal atom and molecular hydrogen shows that H$_2$ binds weakly to Ca-porphyrin 
through a weak electric polarization in dihydrogen, but is strongly hybridized with Ti-porphyrin through the Kubas interaction. 
The presence of $d$ orbitals in Ti may explain the difference in the interaction types.
\end{abstract}


\maketitle

\section{Introduction}

Since hydrogen is a clean and recyclable energy carrier producing no pollution, 
there has been a great deal of effort to find good hydrogen storage materials. 
So far, much attention has been paid to low-dimensional nanostructures, since compact and light-element storage materials are required for mobile applications. 
In particular, carbon-based nanostructures such as carbon nanotubes, fullerenes and polymers were expected to be good candidates for hydrogen storage 
because of their large surface areas to adsorb hydrogen.~\cite{dillon,gulseren,chan,tada,leeec,miura,zhao,dag}
However, pure carbon-based structures have a drawback of insufficient binding energy~\cite{miura,dag} near room temperature:
their binding energies are lower than that for good hydrogen storage which should be 0.2 $-$ 0.3 eV to operate at ambient conditions.~\cite{lee2,lochan,hamaed}
One the other hand, carbon-based nanostructures have been studied as promising catalysts for hydrogen release.\cite{berseth,cento,dehouche} 
Berseth {\em et al.}~\cite{berseth} showed that carbon nanostructures may be used as catalysts for hydrogenation/dehydrogenation of 
sodium aluminium hydride (sodium alanate; NaAlH$_2$).

As candidates having aforementioned optimal binding energies among nanostructures, 
metal-decorated materials such as metal-carbon based materials,~\cite{dag,sun1,yildirim1,kimyh,sun2,durgun,yoon,kimg,ataca}
metal-organic frameworks,~\cite{rosi,yildirim2,huang,srepusharawoot,blomqvist}
and metal-decorated polymers~\cite{lee1,lee2,lee3}
were suggested to increase the potential for hydrogen storage.
For $cis$- and $trans$-polyacetylene decorated with Ti atoms,~\cite{lee1}
up to five and four hydrogen molecules binds to a Ti atom, respectively, and the optimal binding energy of H$_2$ molecules is around 0.2 $-$ 0.3 eV. 
Because of their high cohesive energy ($>$ 4 eV), however, transition metals (TMs) such as Ti and V have a strong tendency of clustering. 
Besides they can act as catalysts to destroy the carbon-based frameworks or to dissociate H$_2$. 
Consequently, the hydrogen storage capability is significantly reduced. 
To prevent clustering of metal atoms, researchers have considered alkali-earth metals (AEMs) with lower cohesive energy (1 $-$ 2 eV). 
Recently, Yoon {\em et al.}~\cite{yoon} showed that a coating of C$_{60}$ fullerenes with Ca could be a non-dissociative hydrogen strorage material, 
using {\em ab initio} calculations within generalized gradient approximation (GGA). 
On the other hand, it is known that alkai metal (AM) elements have relatively low cohesive energies ($<$ 1.7 eV), 
which can also avoid clustering effectively.~\cite{huang} Li-covered graphene were investigated to show that each adsorbed Li on graphene adsorbs 
up to four H$_2$ molecules amounting to 12.8 wt\%.~\cite{ataca}
For alkali-metal-decorated organic molecules, Huang {\em et al.} found that Li-doped cases exhibit a higher storage capacity ($>$ 10 wt\%) 
than Na- and K-doped cases with the adsorption energy of $\sim$0.1 $-$ 0.3 eV/H$_2$.~\cite{huang}
Here, it is noted that AM elements are more reactive than AEM elements in ambient conditions.

As a famous group in organic chemistry, porphyrin consists of four pyrrole-like subunits joined together by four methine (=CH-) groups. 
Four nitrogen atoms in a porphyrin molecule can be saturated by a single metallic ion instead of two H atoms.  
In nature, a heme contains a porphyrin coordinated to Fe, and a chlorophyll contains a porphyrin coordinated to Mg. 
Porphyrins are easy to synthesize and become important molecular building blocks that can be used to make several structures in materials chemistry.~\cite{smithenry,suslick,choi}
Recently, many studies focus on the application of various properties of porphyrin.~\cite{ribeiro,nguyentg}
Previously, we also studied energetics of Ca-porphyrin where Ca is incorporated into the central N$_4$ cavity of porphyrin.~\cite{ryou}

In this paper, we present the different origins of dihydrogen binding to AEM- and TM-inserted porphyrin molecules 
on the basis of first-principles electronic structure calculations. 
Ca and Ti were chosen as metal atoms inserted in the central N$_4$ cavity of porphyrin (C$_{20}$H$_{12}$N$_4$). 
First, we compared energetics of both Ca- and Ti-porphyrins. 
Then we investigated interaction between dihydrogen and an AEM (Ca) or a TM (Ti) atom incorporated to porphyrin. 
As a result, we found that dihydrogen binds weakly to Ca atom through a weak polarization of dihydrogen, 
but interacts strongly with Ti atom with the so-called Kubas interaction. 
Finally, we discussed a possibility of porphyrin-based superstructures as hydrogen storage materials.

\section{Calculation Methods}

We carried out first-principles calculations within local denstiy approximation (LDA) and generalized gradient approximation (GGA), 
using ultrasoft pseudopotentials implemented in the Vienna ab initio Simulation Package (VASP).~\cite{kresse,vanderbilt}
The atomic positions were relaxed with residual forces smaller than 0.01 eV/\AA~and the energy cutoff for a plane wave basis set 
was 400 eV.
The vacuum region was about 10 \AA~between porphyrin molecules, which was enough to avoid the image-image interaction.
First, we optimzed the structure of metalloporphyrin such as Ca- and Ti-porphyrin molecules. 
The inserted metal (Ca or Ti) atom is located at the center of the cavity surounded by four N atoms. 
The Ca atom is placed above the porphyrin plane while Ti atom is located at the same plane to the z-axis in the metalloporphyrin. 
For the L\"owdin charge analyses and the wave function plots, we employed the Quantum Espresso package.

\section{Results and Discussion}

\begin{figure}[b]
\centering \vspace{5mm}
\includegraphics[width=12.4cm]{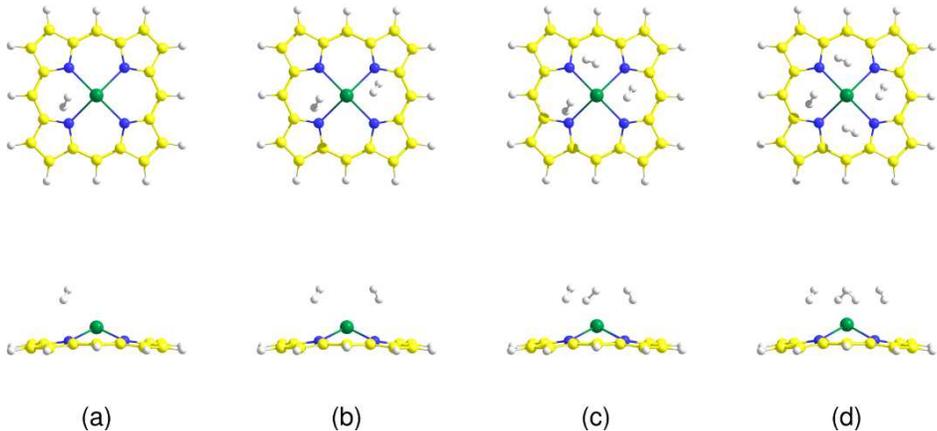}
\caption{Optimized configurations of H$_2$ molecules adsorbed on Ca-porphyrin. The upper and lower panels show top and side views of optimized structures, respectively.
Yellow, blue, green, and white balls represent carbon, nitrogen, calcium, and hydrogen atoms, respectively.} \label{fig1:h2-ca_por} \vspace{5mm}
\end{figure}

\begin{figure}[t]
\centering \vspace{5mm}
\includegraphics[width=8.4cm]{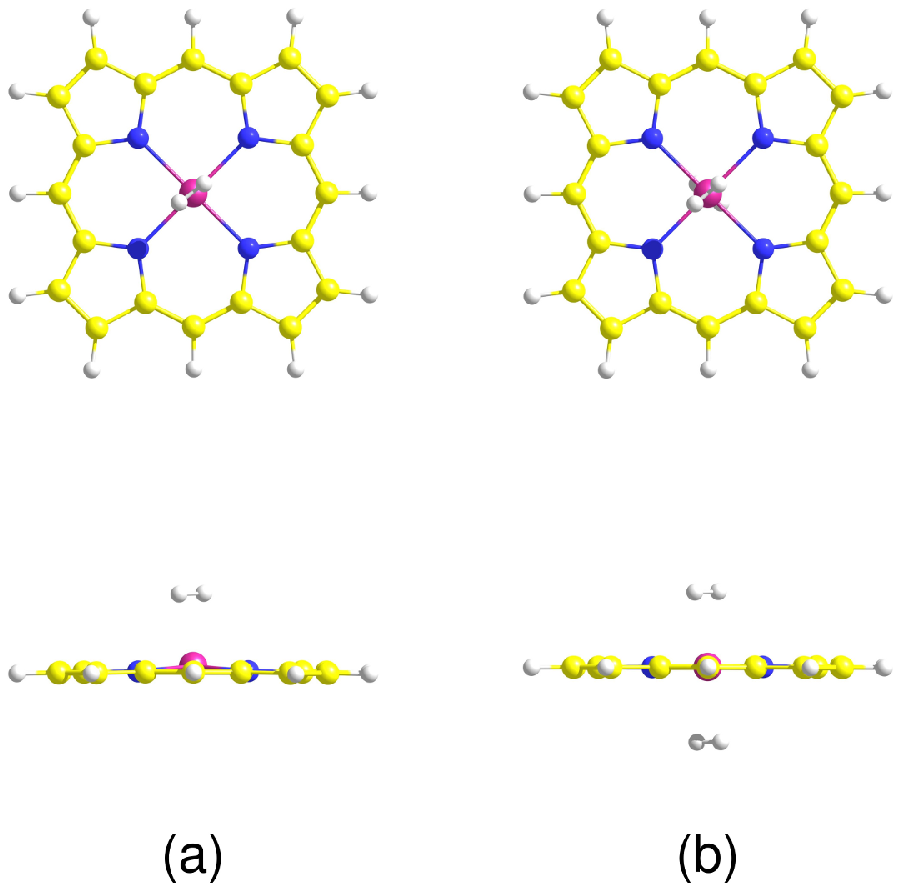}
\caption{Optimized configurations of H$_2$ molecules adsorbed on Ti-porphyrin. 
The upper and lower panels show top and side views of optimized structures, respectively.
Yellow, blue, red-violet, and white balls represent carbon, nitrogen, titanium, and hydrogen atoms, respectively.} \label{fig2:h2-ti_por} \vspace{5mm}
\end{figure}

In a previous paper,~\cite{ryou} we briefly studied H$_2$ adsorption on Ca-porphyrin to find that binding energy of H$_2$ is 
about 0.25 eV per H$_2$, using LDA up to four adsorbed hydrogen molecules. 
Here, we present the detailed results of binding energies and configurations for $n$H$_2$ ($n$ = 1, 2, $\cdots$, 6) 
on Ca-porphyrin along with the case of Ti-porphyrin. The binding energy $E_b$ per H$_2$ for $n$ hydrogen molecules adsorbed on metalloporphyrin (M-porphyrin) is defined by
\begin{equation*}
          E_b = (1/n) ¡¿ [E(\text{M-porphyrin}) + nE(\text{H}_2) - E (n\text{H}_2/\text{M-porphyrin})],
\end{equation*}
where $E$(M-porphyrin), $E$(H$_2$), and $E$($n$H$_2$/M-porphyrin) are the total energies of M-porphyrin, H$_2$, and $n$H$_2$/M-porphyrin, respectively. 
The binding energies for Ca-porphyrin as well as Ti-porphyrin using both LDA and GGA are listed in Table I, 
while the respective optimized geometries are shown in Figs.~\ref{fig1:h2-ca_por} and \ref{fig2:h2-ti_por}.

For Ca-porphyrin, tilted configurations of H$_2$ molecules are the most stable up to 4 adsorbed hydrogen molecules. 
The GGA binding energies are quite small (about 0.1 eV/H$_2$) compared to the LDA values (about 0.25 eV/H$_2$ up to four H$_2$). 
Note that CaC$_{60}$ attracts up to 5H$_2$ molecules with a uniform binding energy of $\sim$0.2 eV/H$_2$ for GGA and $\sim$0.4 eV/H$_2$ for LDA.~\cite{yoon}
Since the binding energies for the fifth and sixth molecules are relatively small, the H$_2$ molecules are considered to be physisorbed on porphyrin. 
For our LDA results of Ca-porphyrin, the bond lengths of H$_2$ molecules were slightly elongated to $\sim$0.78~\AA\ from 0.765~\AA\ (by $\sim$2\%), 
and the distances between H and Ca atoms were 2.35 $-$ 2.52~\AA, which are very similar to the sum of atomic radii of H (0.53~\AA) and Ca (1.94~\AA). 
It means that there is no practical covalent bonding between H and Ca.

Compared with results of Ca-porphyrin, the energetics of H$_2$ adsorption on Ti-porphyrin was also studied. 
The LDA and GGA results for the H$_2$ binding energy on Ti-porphyrin are summarized in Table I. 
Although the number of H$_2$ adsorbed on porphyrin are only two, the binding energy is even larger than that for Ca-porphyrin. 
The LDA (GGA) binding energies per H$_2$ are 0.63 eV (0.32 eV) up to two H$_2$.

\begin{table*}[b]
\caption{The binding energies (eV per H$_2$ molecule) of $n$H$_2$ on metalloporphyrin.}
\label{tab1:bindene}
\begin{ruledtabular}
\begin{tabular}{c|cccccc|cc}
$n$H$_2$/M-porphyrin &\multicolumn{6}{c|}{Ca-porphyrin} &
\multicolumn{2}{c}{Ti-porphyrin}\\
 & 1H$_2$ & 2H$_2$ &3H$_2$ &4H$_2$ &5H$_2$ &6H$_2$ & 1H$_2$ & 2H$_2$ \\ \hline
LDA (eV/H$_2$) & 0.26 &0.26&	0.26&	0.25&	0.21&	0.19&	0.76&	0.63 \\
GGA (eV/H$_2$) & 0.13 &0.12&	0.10&	0.09&	0.07&	0.07&	0.43&	0.32 \\
\end{tabular}
\end{ruledtabular}
\end{table*}

\begin{figure}[b]
\centering \vspace{5mm}
\includegraphics[width=8.4cm]{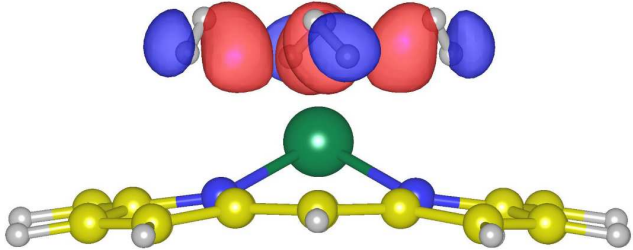}
\caption{Isosurface plot of charge difference between the total electron density of 4H$_2$/Ca-porphyrin and the sum of those of isolated 4H$_2$ and Ca-porphyrin. 
Charge accumulation and depletion are represented by red and blue, respectively. 
The isosurface levels are $\pm$0.0025 electrons/bohr$^3$. A slight electric displacement of hydrogen molecules is shown.} \label{fig3:isos-ca_por} \vspace{5mm}
\end{figure}

Next, we considered a binding mechanism between adsorbed H$_2$ molecules and the Ca atom. 
Though there are several energy levels aligned at the same positions in the PDOS of H$_2$/Ca-porphyrin (Fig.~\ref{fig3:isos-ca_por} in Ref.~\onlinecite{ryou})
as well as 4H$_2$/Ca-porphyrin, no significant overlap (i.e., orbital hybridization) was found between $d$ orbitals of Ca and $\sigma$ orbitals of H$_2$ in the wave functions. 
We calculated the charge difference, $\Delta \rho$, between the configrations before and after adsorption of a hydrogen molecule;

\begin{equation}
\Delta \rho=\rho_{_{\text{nH$_2$/M-porphyrin}}} - \rho_{_{\text{nH$_2$}}} - \rho_{_{\text{M-porphyrin}}}.
\end{equation}

From the L\"owdin charge difference between total charge density of 4H$_2$/Ca-porphyrin and the sum of those of Ca-porphyrin and 4H$_2$, as shown in Fig. 3, 
we found that there is small electric polarization in each hydrogen molecule, 
which is associated with a weak interaction between H$_2$ molecules and the Ca ion. According to our charge analysis, 
Ca donates about 0.8 $e$ to porphyrin, and becomes a positive ion (Ca$^{+}$). 
Consequently, the tilting shape gives rise to the weak electric polarization in a H$_2$ molecule under the electric field by the Ca ion. 
The average difference in charge in the two H atom in H$_2$ is 0.015 $e$. 
It is hard to assign the number of electrons clearly to a specific atom using the L\"owdin charge analysis as well as other charge analysis methods. 
Thus, we have to accept the value of charge with caution. 
This phenomenon was also reported in H$_2$ binding to a Ca-hydroxyl group complex.~\cite{nguyenmc}

\begin{figure}[t]
\centering \vspace{5mm}
\includegraphics[width=8.4cm]{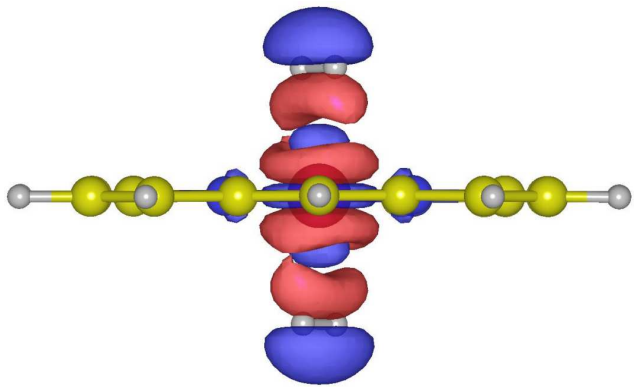}
\caption{Isosurface plot of charge difference between the total electron density of 2H$_2$/Ti-porphyrin and the sum of those of isolated 2H$_2$ and Ti-porphyrin. 
Charge accumulation and depletion are represented by red and blue, respectively. 
The isosurface levels are $\pm$0.0025 electrons/bohr$^3$. Complicated and relatively large electric polarization is shown around Ti and H$_2$ molecules.} 
\label{fig4:isos-ti_por} \vspace{5mm}
\end{figure}

Then, we examined interaction between H$_2$ and Ti-porphyrin. Similar to Ca-porphyrin, we considered a charge difference 
between the total electron density of 2H$_2$/Ti-porphyrin and the sum of those of two isolated H$_2$ and Ti-porphyrin. 
As shown in the charge difference plot in Fig.~\ref{fig4:isos-ti_por},
an electric displacement around both Ti and two H$_2$ occurs along the vertical line to the porphyrin. 
Such relatively large and complicated electric displacement is related to the Kubas interaction, 
which is roughly defined by a sort of interaction between $d$ orbitals of metal atom and $\sigma$ (or $\sigma^*$) orbitals of H$_2$.~\cite{kubas}
In addition, it shows that charge redistribution takes place in the Ti orbitals (especially for $d_{z^2}$) because of adsorption of H$_2$. 
For the case of Ti-porphyrin using LDA, the bond lengths of H$_2$ were considerably elongated to 0.796 $-$ 0.810~\AA\ from 0.765~\AA\ (by $\sim$ 4 $-$ 6\%), 
and the distances between H and Ti atoms were 1.9 $-$ 2.0~\AA, which are shorter than the sum of atomic radii of H (0.53~\AA) and Ti (1.76~\AA). 
It implies that H$_2$-Ti binding is stronger than H$_2$-Ca binding.

\begin{figure}[b]
\centering \vspace{5mm}
\includegraphics[width=14.4cm]{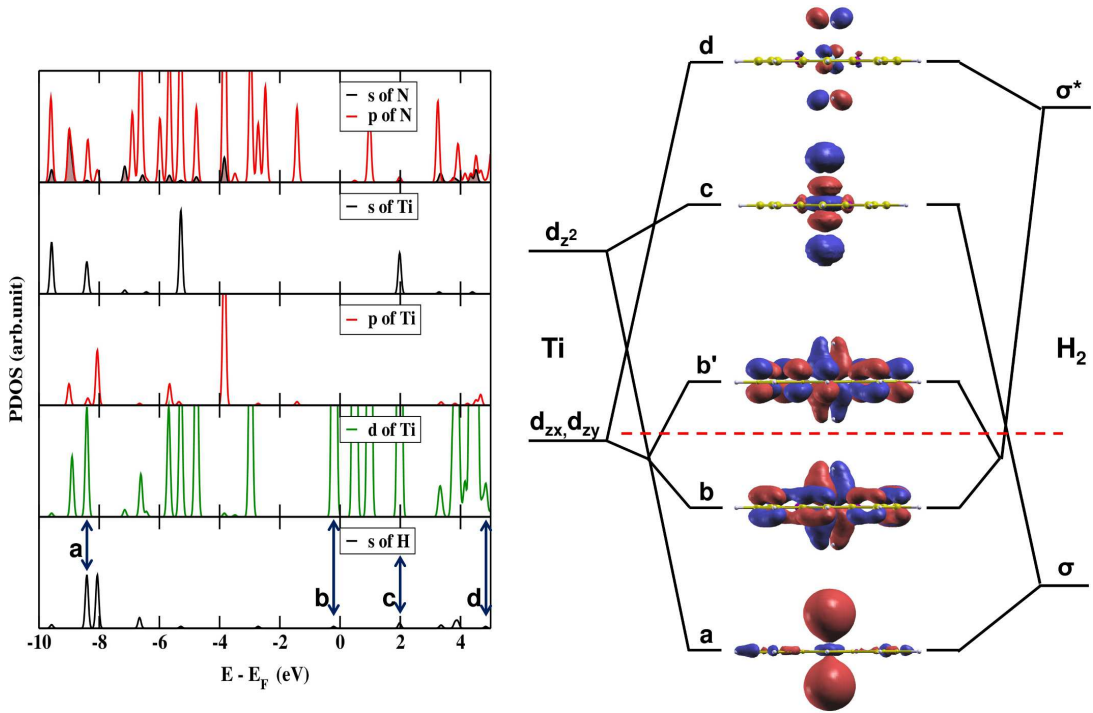}
\caption{(a) PDOS and (b) hybridization diagram for 2H$_2$/Ti-porphyrin. The dotted line in the right panel represents the Fermi level.} \label{fig5:pdos+hydrid} \vspace{5mm}
\end{figure}

To understand interaction between H$_2$ and Ti atom in more detail, we considered the electronic properties of the 2H$_2$/Ti-porphyrin system 
such as the projected density of states (PDOS) and wave function characters. 
The PDOS of Ti, adsorbed H$_2$ molecules and surrounding N atoms is shown in Fig.~\ref{fig5:pdos+hydrid}(a), 
where some (apparent) hybridization are seen between $d$ orbitals of Ti and $s$ orbital of H (or N). 
After detailed investigations of wave function characters for the coincident energy levels, 
we obtained a hybridization diagram in Fig.~\ref{fig5:pdos+hydrid}(b), 
where we found the bonding and anti-bonding states between Ti 3$d$ orbitals and $\sigma$ (or $\sigma^*$) orbitals of H$_2$. 
Each wave function character labeled `$a$'-`$d$' corresponds to each denoted electronic states having the same label in Fig.~\ref{fig5:pdos+hydrid}(a).~\cite{phase}

Three hybridizations labeled `$a$'-`$c$' are clearly noticeable. 
The first one denoted by `$a$' is a hybridized bonding state of the empty $d_{z^2}$ orbital of Ti 
with filled $\sigma$ orbitals of H$_2$, where the filled $\sigma$ orbital donates electrons to empty $d_{z^2}$ orbital. 
The second one denoted by `$b$' just below the Fermi energy is another hybridized bonding state of the filled $d$ orbitals ($d_{zx}$, $d_{zy}$) 
of Ti with empty $\sigma^*$ orbitals of H$_2$, where the filled $d$ orbitals ($d_{zx}$, $d_{zy}$) back-donate electrons to empty $\sigma^*$ orbitals. 
This bonding state has a `twin' anti-bonding state denoted by `$b$'  which is located just above the Fermi level: 
Due to surrounding environments around Ti-H$_2$, these two states ($b$ and $b'$) would be split from an original bonding state hybridized 
from the filled d orbitals ($d_{zx}$, $d_{zy}$) of Ti with empty $\sigma^*$ orbitals of H$_2$. 
On the other hand, we have a distinct anti-bonding state (denoted by `$c$') obtained from the $d_{z^2}-\sigma$ hybridization. 
Another anti-bonding state between $d$ orbitals ($d_{zx}$, $d_{zy}$) and $\sigma^*$ of H$_2$ is not clearly shown. 
As a result, we conclude that H$_2$ binds strongly to Ti atom through the Kubas interaction. 
It should be compared with a weak interaction between H$_2$ and Ca through slight polarization of hydrogen molecules. 
A difference in the interaction types between Ti- and Ca-porphyrin molecules may be attributed to the presence of $d$ orbital in Ti.

Finally, we turned our interest to a possibility of metalloporphyrin as a hydrogen storage material by calculating the hydrogen storage capacity. 
When hydrogen molecules are directly adsorbed on the Ca atom, the storage capacity is estimated to about 2.3 wt\% for 4H$_2$/Ca-porphyrin. 
On the other hand, the hydrogen storage capacity of Ti-porphyrin is about 1.1 wt\% for 2H$_2$/Ti-porphyrin. 
Unfortunately, these values are much smaller than the requirement for practical applications of hydrogen storage. 
Whereas model systems of other research groups are carbon-based structures with metal adatom, our model has strong covalent bonding between the Ca (or Ti) atom 
and four nitrogen atoms in porphyrin. Thus, there are not so reactive electronic orbitals to interact with H$_2$ in metalloporphyrin. 
Consequently, the storage capacity of metal-incorporated porphyrin is not as high as those of metal-decorated carbon nanostrcutures. 
If metal organic frameworks (MOFs) are constructed using metal-incoporated porphyrins,~\cite{bezzu}
the storage capacity could increase even in ambient conditions. Therefore, further investigations are needed for metal-incorporated porphyrin molecules 
to seek high-capacity H$_2$ storage materials. 
At present, gravimetric storage capacities of MOFs are much smaller than 1 wt\% at ambient pressures and room temperature.~\cite{li1,li2}

\section{Conclusion}

In summary, we have investigated non-dissociative molecular hydrogen binding onto M-inserted (M = Ca, Ti) porphyrin, 
and have compared the differences in the origins of dihydrogen binding to Ca- and Ti-porphyrin, using density functional theory calculations. 
By successively increasing the number of adsorbed hydrogen molecules on a Ca-porphyrin or Ti-porphyrin, 
we obtained the optimized structures and the H$_2$ binding energies. 
According to our LDA (GGA) calculations, the binding energy of H$_2$ to the Ca atom is about 0.25 (0.1) eV/H$_2$ 
up to four H$_2$ molecules while that to the Ti atom is about 0.6 (0.3) eV/H$_2$ up to two H$_2$. Our results demonstrate 
that a calcium atom incorporated into porphyrin with H$_2$ molecules forms a dihydrogen complex mainly through a weak polarization of H$_2$. 
In contrast, Ti-porphyrin generates a Kubas-type dihydrogen complex with H$_2$. 
The presence of $d$ orbital in the Ti element may explain the difference in the interaction types.

\begin{acknowledgments}
This research was supported by the Converging Research Center
Program (2010K001069), Priority Research Centers Program
(2010-0020207) and the Basic Science Research Program
(KRF-2008-313-C00217) through the National Research Foundation of
Korea (NRF) funded by the Ministry of Education, Science and
Technology(MEST). Calculations were performed by using the
supercomputing resources of KISTI. 
G. K. acknowledges valuable discussion with Moon-Hyun Cha and support by the Basic Science Research Program through MEST/NRF (2010-0007805).
\end{acknowledgments}


\begin{thebibliography}{99}
\bibitem{phase}
For phase information, wave function characters were plotted using the Quantum Espresso package.
\end{thebibliography}
\end{document}